\documentclass[12pt]{article}
\usepackage{amssymb}
\usepackage{epsfig}
\usepackage{graphics}
\usepackage{graphicx}
\usepackage{color}

\usepackage[latin1]{inputenc}
\usepackage[english]{babel}

\usepackage{amssymb}
\usepackage{amsmath}
\usepackage{amsfonts}
\usepackage{bbm}
\usepackage{booktabs}
\usepackage{ctable}
\usepackage{subfigure}
\usepackage{float}
\usepackage{layout}
\usepackage{fancyhdr}
\usepackage{color}
\usepackage{eurosym}
\usepackage{wrapfig}  

\usepackage{xr}

\def\cvd{$\quad\square$\bigskip}

\def\<{\langle}
\def\>{\rangle}

\newtheorem{theorem}{Theorem}

\newtheorem{proposition}[theorem]{Proposition}
\newtheorem{remark}[theorem]{Remark}

\definecolor{mycolorgreen}{cmyk}{1, .1, 1, .4}

\definecolor{mycolororange}{cmyk}{0.05, 0.5, 1, 0}

\definecolor{mycolormagenta}{cmyk}{0, 1, 0, 0}

\definecolor{mycolorpurple}{cmyk}{0.45, 0.86, 0, 0}

\definecolor{mycolorblue}{rgb}{0, 0, 1}

\definecolor{mycolorred}{rgb}{1, 0, 0}

\definecolor{mycoloryellow}{cmyk}{0, 0, 1, 0}

\definecolor{mycolorblack}{cmyk}{0, 0, 0, 1}

\date{}

\begin{document}
\parindent 0pt

\title{
\bf
Efficient tree methods\\
for pricing digital barrier options\\
}
\author{
{\sc Elisa Appolloni}\\
\small{Sapienza Universit\`a di Roma}\\
\small{MEMOTEF}\\
\small{{\tt elisa.appolloni@uniroma1.it}}
\and
{\sc Andrea Ligori}\\
\small{Universit\`a di Roma Tor Vergata}\\
\small{Dip.to di Matematica}\\
\small{{\tt andrea.ligori@live.it}}
}

%
\maketitle
\begin{abstract}\noindent{\parindent0pt
We propose an efficient lattice procedure which permits to obtain European and American option prices under the Black and Scholes model for digital options with barrier features. Numerical results show the accuracy of the proposed method.
}
\end{abstract}


\smallskip

\section{Introduction}

Tree-based algorithms for option pricing are studied since the seminal work of \textcolor{blue}{Cox \textit{et al.}} (\textcolor{blue}{1979}) and turn out to be very simple and fast to be implemented by a backward induction. An important characteristic which makes these procedures very appealing is that they easily include American-style features once the European case is treated and well setup. This makes lattice techniques widely used in the practice because although many progresses have been done in the development of exact formulas or other numerical procedures (Monte Carlo, finite differences, etc.) for European option prices, the American counterparts, that involve a control problem, are not so well-provided.
We consider here the Black and Scholes model, see \textcolor{blue}{Black\&Scholes} (\textcolor{blue}{1973}) and \textcolor{blue}{Merton} (\textcolor{blue}{1973}), which is either classical and still widely used in finance. It means that the underlying asset price process evolves as a geometric Brownian motion.
Here, option prices can be computed by using the simple tree method due to \textcolor{blue}{Cox \textit{et al.}} (\textcolor{blue}{1979}) (CRR). However financial derivatives have been becoming more and more sophisticated and this means that the standard implementation of the CRR binomial tree brings to further errors in the approximation of the Black and Scholes prices. This is the reason why it becomes important to setup ``efficient tree schemes'', i.e. tree methods which allow one to reduce the approximation errors.\\
\mbox{\quad} We propose here a lattice scheme for pricing digital barrier options. In particular a digital call option is an option whose payoff is equal to a fixed amount (in what follows we suppose this amount is equal to $1$) if the underlying asset at maturity is greater than a predetermined level (the strike price $K$) or nothing otherwise. Practitioners that trade these products
essentially predict the direction of the market without concerning in the specific the magnitude of the movements of the underlying asset price. One of the benefits with respect to standard products is that the investment and the returns are fixed, so the risk involved and the potential losses are known a priori.
Digital options can also include barrier levels: they can be activated or nullified if the underlying asset price process reaches certain contractually specified levels. This more complex option can be used as a financial tool embedded in sophisticated products, such as accrual range notes. These notes are financial securities that are linked for example to a foreign exchange rate and then they pay a fixed interest accrual if the exchange rate remains within a specified range and nothing otherwise, see \textcolor{blue}{Wystup} (\textcolor{blue}{2006}) and also \textcolor{blue}{Hui} (\textcolor{blue}{1996}).\\
\mbox{\quad}European digital options with barrier features have a known closed-form pricing formula both for a single barrier and for double barriers that can be easily derived from the corresponding formulas in the standard case, for the latter ones we can refer for instance to \textcolor{blue}{Reiner\&Rubinstein} (\textcolor{blue}{1991}) in the single barrier case and \textcolor{blue}{Ikeda\&Kunitomo} (\textcolor{blue}{1992}) in the double barrier case. As far as we know, the literature of American exotic options suggests no approximation formulas for digital barrier options. Our objective is then to treat the option pricing problem related to these options by using lattice techniques. In particular we propose an efficient method for the pricing of European digital call options with a single barrier and then, consequently, we also get a good method for the case of American-style options.\\
\mbox{\quad} To this end, we first need to find an explicit asymptotic expansion of the classical CRR binomial approximation error, that is the difference between the price computed with the CRR tree and the Black and Scholes price. Then we setup an algorithm such that it behaves ``well'', in the sense that the worst contribution in the asymptotic expansion (which turns out to be of order $\frac{1}{\sqrt{n}}$, where $n \in \mathbb{N}$ is the number of time steps of the tree) is nullified.\\
\mbox{\quad} It is well known that the rate of convergence of the CRR tree for vanilla options on continuous payoff functions is of order $\frac{1}{n}$, see for example \textcolor{blue}{Walsh\&Walsh} (\textcolor{blue}{2002}), \textcolor{blue}{Diener\&Diener} (\textcolor{blue}{2004}) and \textcolor{blue}{Chang\&Palmer} (\textcolor{blue}{2007}).
Moreover the results known from the literature when dealing with barrier options always require the continuity on the payoff function. For double barrier options on a general class of continuous payoff functions we can refer to \textcolor{blue}{Gobet} (\textcolor{blue}{2001}) and in the more specific case of call options to \textcolor{blue}{Lin\&Palmer} (\textcolor{blue}{2013}). 
We recall that here the rate of convergence is of order $\frac{1}{\sqrt{n}}$ and this is due to the fact that the contractual barriers do not necessarily coincide with the effective barriers on the tree structure.
On the other hand when the payoff is assumed to be discontinuous the analysis of the rate of convergence of the binomial algorithm is given only for vanilla options, see \textcolor{blue}{Walsh\&Walsh} (\textcolor{blue}{2002}) and \textcolor{blue}{Chang\&Palmer} (\textcolor{blue}{2007}).
Also in this case the rate of convergence is of order $\frac{1}{\sqrt{n}}$, but now it is caused by the position of the discontinuity points of the payoff functions on the lattice.\\
\mbox{\quad }In this framework we treat the study of the CRR binomial approximation error in the case of digital options with a single barrier and we get a complete theoretical result that allows us to construct an efficient algorithm. In particular we fit the Binomial Interpolated Lattice (BIL) algorithm provided in \textcolor{blue}{Appolloni \textit{et al.}} (\textcolor{blue}{2013}) to this specific case. We recall that the BIL procedure is based on a backward induction on a binomial mesh with additionally suitable interpolations that allows one to get very precise option prices for call and put double barrier options. We adapt this procedure to the case of digital call and put options with a single barrier and the numerical results show that we get very reliable and accurate option prices.\\
\mbox{\quad} The paper is organized as follows. In Section \ref{sect-model} we describe the continuous-time model for the evolution of the underlying asset price process. In Section \ref{sect-theory} we propose our theoretical contribution on the asymptotic expansion of the CRR approximation error for European digital options with barrier features. The numerical results are shown in Section \ref{sect-numerics}.

\section{The model}\label{sect-model}
We consider a market model in the time interval $[0,T]$ where the evolution of the risky asset $(S_t)_{t \in [0, T]}$ is governed by the Black-Scholes stochastic differential
equation
\begin{equation}\label{sde} 
\frac{dS_t}{S_t}=rdt+\sigma dB_t ,  \quad S_0=s_0 > 0, \quad
\end{equation}
where $(B_t)_{t \in [0, T]}$ is a standard Brownian motion under the risk neutral probability measure. The non-negative constant $r$ is the risk-free interest rate and $\sigma$ is the constant volatility parameter.\\
\mbox{\quad}A digital call barrier option is a contract whose payoff is equal to $1$ if the underlying asset price at maturity is greater than the strike price $K$, and nothing otherwise. Moreover, the payoff also depends on whether the underlying stock price path ever touches certain price levels called barriers. Once either of these barriers is breached, the status of the option is immediately determined: either the option comes into existence if the barrier is a knock-and-in type or it ceases to exist if the barrier is a knock-and-out type. In what follows we will consider the knock-and-out type, the knock-and-in type being similar.\\
We recall that the payoff of a digital call option with lower barrier $L$ is given by
\begin{equation}\label{payoff-single}
\mathbbm{1}_{S_T \geq K}\mathbbm{1}_{S_{\inf}>L},
\end{equation} 
where $S_{\inf}=\inf_{t \in [0,T]} S_t$. The case of a digital call option with higher barrier $H$ is similar, as well as the case of the corresponding digital put options with a single barrier.\\
\mbox{\quad}The Black and Scholes prices of European options whose payoff is equal to the one given in (\ref{payoff-single}) can be found in \textcolor{blue}{Reiner\&Rubinstein} (\textcolor{blue}{1991}). No analytical approximations are available for option prices when the payoff is as in (\ref{payoff-single}) in the American case.

\section{The CRR binomial approximation error}\label{sect-theory}

Starting from the model described in Section \ref{sect-model}, we analyze here the error committed by using the CRR tree method for pricing European digital call options with a single barrier. If $n \in \mathbb{N}$ denotes the number of time steps of the tree, we define the CRR binomial approximation error as follows
\begin{equation}\label{error}
\mbox{Err}_{CRR}(n)= \mbox{price}_{CRR}(n) - \mbox{price}_{BS}, \quad \forall \, n \in \mathbb{N}
\end{equation}
where $\mbox{price}_{CRR}(n)$ denotes the price calculated by using the CRR tree scheme and $\mbox{price}_{BS}$ is the Black and Scholes price.\\
\mbox{\quad}We briefly recall here the idea of the CRR tree scheme. Let us fix an integer $n \in \mathbb{N}$. We first define $\Delta \tau=T/n$, where $T$ is the maturity of the option, and then we build a binomial tree with $n$ steps of length $\Delta \tau$. If we label $(0,0)$ the starting node that corresponds to the value $S_0=s_0$, then after $i$ time steps ($i=0,1,...,n$), the discrete process may be located at one of the nodes $(i,j)$ ($j=0,1,...,i$) corresponding to the values
\begin{equation}
S_{i,j}=s_0e^{(2j-i)\sigma\sqrt{\Delta \tau}}.
\end{equation} 
Hence, starting from $S_{i,j}$ at time $ih$, the process may jump at time $(i+1)h$ to the value $S_{i+1,j+1}$ or the value $S_{i+1,j}$ with probability $p$ and $1-p$ respectively, where $p$ is defined as
\begin{equation}\label{proba}
p=\frac{e^{r\Delta \tau}-d}{u-d}
\end{equation}
and $u=e^{\sigma\sqrt{\Delta \tau}}=d^{-1}$.
European and American prices at time $0$ are then obtained by applying the classical backward induction.\\
\mbox{\quad} 

\mbox{\quad} We now develop some argumentations described in \textcolor{blue}{Chang\&Palmer} (\textcolor{blue}{2007}) and \textcolor{blue}{Lin\&Palmer} (\textcolor{blue}{2013}). We consider the case of a digital call option with a lower barrier $L$, the other cases of digital options with a single barrier being similar. The idea is to find a closed-form formula in terms of binomial coefficients of the option price following \textcolor{blue}{Reimer\&Sandmann} (\textcolor{blue}{1995}) and then use the approximation of the binomial distribution by the normal one as suggested in \textcolor{blue}{Lin\&Palmer} (\textcolor{blue}{2013}) in order to find explicit coefficients in the asymptotic expansion.

\mbox{\quad}We first stress that we need the binomial prices of digital call options with lower barrier $L$ of the two following types:
\begin{enumerate}
\item
a down-and-in call option with $L<K$;
\item
a down-and-out call option with $L>K$.
\end{enumerate}
In these two cases the binomial formulas are manageable and permits a simple treatment. Then, by using the binomial formulas for the corresponding vanilla digital call option, it is possible to find the asymptotic expansion of the error for a down-and-out digital call option with $L<K$ and a down-and-in digital call option with $L>K$.\\

\mbox{\quad} Let us first introduce two quantities, that will have a crucial role in what follows, as defined in \textcolor{blue}{Chang\&Palmer} (\textcolor{blue}{2007}) and \textcolor{blue}{Lin\&Palmer} (\textcolor{blue}{2013}), that we call $\Delta^K_n$ and $\Delta^L_n$.
The quantity $\Delta^K_n$ is set as
\begin{equation}\label{dkn}
\Delta^K_n=1-2\mbox{frac}\Biggl(\frac{\log(s_0/K)}{2\sigma\sqrt{\Delta \tau}}-\frac{n}{2}\Biggl),
\end{equation}
where for every real number $x$, the fractional part of $x$ is defined as $\mbox{frac}(x)=x-\lfloor x \rfloor$, with $\lfloor x \rfloor$ indicating the largest integer preceding $x$. We observe that $\Delta^K_n$ is a measure of the position of $K$ in the $\log$-scale in relation to two adjacent terminal stock prices. In fact if we define $j_K$ the integer such that
$$
S_{n,j_K-1}=s_0u^{j_K-1}d^{n-j_K+1} < K \leq S_{n,j_K}=s_0u^{j_K}d^{n-j_K},
$$ 
then $\Delta^K_n=-1$ if $\log K$ is the node $\log S_{n,j_k-1}$ at maturity, $\Delta^K_n=0$ if $\log K$ lies halfway between the two nodes $\log S_{n,j_k-1}$ and $\log S_{n,j_k}$ at maturity (i.e. it is a node from the first period before maturity) and $\Delta^K_n=1$ if $\log K$ is the node $\log S_{n,j_k}$ at maturity.\\
We now describe a \textit{similar} quantity corresponding to the lower barrier $L$ that we call $\Delta^L_n$. First, we call $\tilde{L}$ the effective barrier on the tree structure, that is generally different from the contractual barrier $L$. Let us suppose that $j_L$ is the number of up jumps required to reach $\tilde{L}$. We define
$$
\Delta^L_n =\mbox{frac}(2l_L), \quad \mbox{with} \quad l_L=\frac{\log \frac{L}{s_0}}{2\sigma\sqrt{\Delta \tau}}+\frac{n}{2}.
$$
Then the effective barrier $\tilde{L}$ can be written as $\tilde{L}=s_0u^{\tilde{j}_L}d^{n-\tilde{j}_L}$, where 
$$
\tilde{j}_L = j_L+\frac{1}{2}(1-\epsilon_n),
$$
with
$$
j_L=\frac{1}{2}\lfloor 2l_L \rfloor
$$
and 
$$
\epsilon_n = \bigg \{
\begin{array}{rl}
&0, \quad \mbox{if the effective barrier is not a terminal stock price},\\
&1, \quad \mbox{if the effective barrier is a terminal stock price}. \\
\end{array}
$$

The quantity $\Delta^L_n \in [0,1]$ measures in the $\log$-scale the position of $L$ in relation to two adjacent stock prices, one of which is a node at maturity and the other is a node of the first time before maturity. In the special cases in which $\Delta^L_n=0$ and $\Delta^L_n=1$ we get that the effective barrier $\tilde{L}$ lies exactly on a node of the tree (for a more detailed discussion on this see \textcolor{blue}{Lin\&Palmer} (\textcolor{blue}{2013})).\\

\mbox{\quad}We now need to introduce some notations as in \textcolor{blue}{Reimer\&Sandmann} (\textcolor{blue}{1995}). We define $\pi_d(n,j,\tilde{j}_L)$ as the price at time $0$ of a security which pays on unit at time $T$ if the asset price at the time step $n$ is equal to $S_{n,j}=s_0u^jd^{n-j}$ and if there exists a pair $(i,l)$ with $i \in \{0,...,n\}$ and $l \in \{0,...i\}$, such that $S_{i,l}=s_0u^ld^{i-l} \leq \tilde{L}$, and otherwise nothing, i.e.
\begin{align}\label{p-bino}
\pi_d(n,j,\tilde{j}_L)&=e^{-rT}\mathbb{E}[\mathbbm{1}_{S^n_{T}=S_{n,j}}\cdot\mathbbm{1}_{\exists \, i\leq n,\exists \,l \leq i:S_{i,l}=s_0u^ld^{i-l}\leq \tilde{L}}]\notag\\
&=e^{-rT}\mathbb{P}(S^n_{T}=S_{n,j};\exists \, i\leq n,\exists \, l \leq i:S_{i,l}\leq \tilde{L}),\notag\\
\end{align}
where $(S^n_{t_i})_{i=0,1,...,n}$, with $t_i=i\Delta \tau$ for every $i=0,1,...,n$, denotes the discrete approximation of $S_{ih}$, so in particular $S^n_{t_n}=S^n_T$ is the discrete approximation of $S_T$.
In order to calculate (\ref{p-bino}), we first need to count the number of paths $Z_d(n,j,\tilde{j}_L)$ in the binomial tree which reach the terminal stock price $S_{n,j}$ after touching or passing through the effective barrier $\tilde{L}$. The reflection principle (see \textcolor{blue}{Feller} (\textcolor{blue}{1968})) yields the number $Z_d(n,j,\tilde{j}_L)$ that for every $j=0,...,n$ is equal to
\begin{equation}\label{sistema-path}
Z_d(n,j,\tilde{j}_L)=\left\{
\begin{array}{l}
\binom{n}{j}, \quad \mbox{if} \quad j\leq \tilde{j}_L, \\
\binom{n}{2\tilde{j}_L-j}, \quad \mbox{if} \quad \tilde{j}_L < j \leq 2\tilde{j}_L, \\
0, \quad \mbox{if} \quad j>2\tilde{j}_L.\\\end{array}
\right.
\end{equation} 

We can now prove the following Proposition:
\begin{proposition}\label{prop-1}
The binomial price $C_{di-digital}(s_0, K, T, L, n)$ of a down-and-in digital call option with barrier $L<K<s_0$ is equal to
\begin{equation}\label{binomial-di}
C_{di-digital}(s_0, K, T, L, n) = e^{-rT}\sum_{i=j_K}^{2\tilde{j}_L}\binom{n}{2\tilde{j}_L-i}p^i(1-p)^{n-i}.
\end{equation}
\end{proposition}

\textit{Proof.}
Let us denote with $G(S_{n,j})$ the payoff of a digital call option at node $S_{n,j}$, i.e.
\begin{equation}\label{payoff-G}
G(S_{n,j})=\left\{
\begin{array}{l}
1, \quad \mbox{if} \quad S_{n,j}\geq K, \\
0, \quad \mbox{if} \quad S_{n,j}<K.\\
\end{array}
\right.
\end{equation}
So the price at time $0$ of a down-and-in digital call option is equal to
\begin{align}
&C_{di-digital}(s_0, K, T, L, n)=e^{-rT}\mathbb{E}[G(S^n_{T})\cdot \mathbbm{1}_{\exists \, i\leq n,\exists \,l \leq i:S_{i,l}=s_0u^ld^{i-l}\leq \tilde{L}}]\notag\\
&=e^{-rT}\sum_{j=0}^n\mathbb{E}[G(S^n_{T})\mathbbm{1}_{S^n_{T}=S_{n,j}}\mathbbm{1}_{\exists \, i\leq n,\exists \,l \leq i:S_{i,l}=s_0u^ld^{i-l}\leq \tilde{L}}]=\sum_{j=j_K}^n\pi_d(n,j,\tilde{j}_L)\notag\\
&=e^{-rT}\sum_{j=j_K}^n\Biggl[\binom{n}{j}p^j(1-p)^{n-j}\mathbbm{1}_{j\leq \tilde{j}_L}+\binom{n}{2\tilde{j}_L-j}p^j(1-p)^{n-j}\mathbbm{1}_{\tilde{j}_L<j\leq2\tilde{j}_L}\Biggl]\notag\\
&=e^{-rT}\sum_{j=j_K+1}^{2\tilde{j}_L}\binom{n}{2\tilde{j}_L-j}p^j(1-p)^{n-j},
\end{align}
where the last equality comes from the fact that here we suppose $L<K$ (i.e. $\tilde{j}_L<j_K$) and as a consequence one has that the contribution due to the first sum vanishes. So the proof is complete.
\cvd

\mbox{\quad}We now derive the price of a down-and-out digital call option with $L>K$, that we call
$C_{do-digital}(s_0, K, T, L, n)$, as the difference between the binomial price of the vanilla digital call option and the binomial price of the down-and-in digital call option with $L>K$.\\
Let us start from the binomial price of the vanilla digital call option, that we denote with $C_{digital}(s_0,K,T,n)$. Let us call with $\pi(n,j)$ the price at time $0$ of a security which pays one unit at time $T$ if the asset price is equal to $S_{n,j}=s_0u^jd^{n-j}$ and otherwise nothing, i.e.
$$
\pi(n,j)=e^{-rT}\mathbb{E}[\mathbbm{1}_{S^n_{T}=S_{n,j}}]=e^{-rT}\binom{n}{j}p^j(1-p)^{n-j}.
$$
We denote as before with $G(S_{n,j})$ the payoff of a digital call option at node $S_{n,j}$, see (\ref{payoff-G}).
So the price at time $0$ of a vanilla digital call option is equal to
\begin{align}\label{vanilla}
&C_{digital}(s_0,K,T,n) = e^{-rT}\mathbb{E}[G(S^n_T)]=e^{-rT}\sum_{j=0}^n\mathbb{E}[G(S^n_T)\mathbbm{1}_{S^n_T=S_{n,j}}]\notag\\
&=\sum_{j=j_K}^n \pi(n,j)=e^{-rT} \Biggl[\sum_{j=j_K}^{n}\binom{n}{j}p^j(1-p)^{n-j}\Biggl].\notag\\
\end{align}
Now from the proof of Proposition \ref{prop-1} we know that the binomial price of a down-and-in digital call option with $L>K$ (i.e. $\tilde{j}_L > j_K$) can be written as
\begin{align}\label{out}
&C_{di-digital}(s_0,K,T,L,n)\notag\\
&=e^{-rT}\Biggl[\sum_{j=j_K}^{\tilde{j}_L}\binom{n}{j}p^j(1-p)^{n-j}+\sum_{j=\tilde{j}_L+1}^{2\tilde{j}_L}\binom{n}{2\tilde{j}_L-i}p^j(1-p)^{n-j}\Biggl],\notag\\
\end{align}
so we can state the following result:
\begin{proposition}\label{prop-2}
The binomial price $C_{do-digital}(s_0, K, T, L, n)$ of a down-and-out digital call option with barrier $L>K$ is equal to
\begin{align*}
C_{do-digital}(s_0, K, T, L, n) &= e^{-rT}\Biggl[\sum_{i=\tilde{j}_L+1}^{n}\binom{n}{i}p^i(1-p)^{n-i}\\
&-\sum_{i=\tilde{j}_L+1}^{2\tilde{j}_L}\binom{n}{2\tilde{j}_L-i}p^i(1-p)^{n-i}\Biggl].
\end{align*}
\end{proposition}

\textit{Proof.}
The price $C_{do-digital}(s_0, K, T, L, n)$ of a down-and-in digital call option with $L>K$ is then obtained by subtracting the price $C_{digital}(s_0,K,T,L)$ given in (\ref{vanilla}) to the price
 $C_{di-digital}(s_0,K,T,L,n)$ given in (\ref{out}).
\cvd

\mbox{\quad}We now give the explicit coefficients of order $\frac{1}{\sqrt{n}}$ and $\frac{1}{n}$ in the asymptotic expansion of the CRR binomial error defined in (\ref{error}) for the price of digital call options with barrier $L$. The idea is to use the closed-form formulas of the binomial prices given in Proposition \ref{prop-1} and Proposition \ref{prop-2} and then approximate them by using Lemma 4.1 in \textcolor{blue}{Lin\&Palmer} (\textcolor{blue}{2013}) on the approximation of the binomial distribution by the normal one.\\
We can state the following result:
\begin{theorem}\label{digital-1}
In the $n$-period CRR binomial model, the binomial error $\mbox{Err}_{CRR}(n)$ for the prices of European digital call options with barrier $L<K$ is:
\begin{itemize}
\item
for a down-and-in digital call option:
\begin{align*}
\mbox{Err}_{CRR}(n)&=e^{-rT}\Biggl[(\tilde{A}_1\Delta^K_n+\tilde{A}_2\Delta^L_n)\frac{1}{\sqrt{n}}\\
&+(\tilde{B}_1+\tilde{B_2}(\Delta^K_n)^2+\tilde{B}_3\Delta^K_n\Delta^L_n+\tilde{B}_4(\Delta^L_n)^2)\frac{1}{n}\Biggl]+O\Biggl(\frac{1}{n^{3/2}}\Biggl);\\
\end{align*}
\item
for a down-and-out digital call option:
\begin{align*}
\mbox{Err}_{CRR}(n)&=e^{-rT}\Biggl[(\tilde{C}_1\Delta^K_n+\tilde{C}_2\Delta^L_n)\frac{1}{\sqrt{n}}\\
&+(\tilde{D}_1+\tilde{D}_2(\Delta^K_n)^2+\tilde{D}_3\Delta^K_n\Delta^L_n+\tilde{D}_4(\Delta^L_n)^2)\frac{1}{n}\Biggl]+O\Biggl(\frac{1}{n^{3/2}}\Biggl).\\
\end{align*}
\end{itemize}
The list of the constant is postponed in Appendix A.
\end{theorem}

\textit{Proof.}
Let us consider first the binomial price of the down-and-in digital call option given in Proposition \ref{prop-1}, that is
\begin{equation}\label{e1}
C_{di-digital}(s_0,K,T,L,n)=e^{-rT}\sum_{i=j_K}^{2\tilde{j}_L}\binom{n}{2\tilde{j}_L-i}p^i(1-p)^{n-i}.
\end{equation}
From equation (5.1) in \textcolor{blue}{Lin\&Palmer} (\textcolor{blue}{2013}) we can write (\ref{e1}) as follows
\begin{equation}\label{e2}
C_{di-digital}(s_0,K,T,L,n)=e^{-rT}\Biggl(\frac{1-p}{p}\Biggl)^{n-2\tilde{j}_L}\sum_{i=0}^{2\tilde{j}_L-j_K}\binom{n}{i}p^{n-i}(1-p)^i.
\end{equation}
The asymptotic expansion of (\ref{e2}) is now obtained by applying the asymptotic expansion (5.7) in \textcolor{blue}{Lin\&Palmer} (\textcolor{blue}{2013}) of the term $(\frac{1-p}{p})^{n-2\tilde{j}_L}$ and the asymptotic expansion (5.3) in \textcolor{blue}{Lin\&Palmer} (\textcolor{blue}{2013}) of the term $\sum_{i=0}^{2\tilde{j}_L-j_K}\binom{n}{i}p^{n-i}(1-p)^i$.\\
The asymptotic expansion for the down-and-out digital call option is now straightforward. In fact it can be derived from the asymptotic expansion for the down-and-in option and that for the corresponding vanilla option that can be found in \textcolor{blue}{Chang\&Palmer} (\textcolor{blue}{2007}).
\cvd

\begin{remark}
Theorem \ref{digital-1} shows that the contribution of the type $\frac{1}{\sqrt{n}}$ in the asymptotic expansion of $\mbox{Err}_{CRR}(n)$ is due to the position of barrier and the position of the strike with respect to the nodes of the tree. In order to obtain an algorithm of order $\frac{1}{n}$, we need to set $\Delta^K_n=0$ and $\Delta^L_n=0$. It means that in the $\log$-scale the strike $K$ must be positioned halfway between two nodes at maturity (i.e. it should be a node of the penultimate period before maturity) and the barrier $L$ must lie on a layer of nodes of the tree.
\end{remark}

We now state the following result:

\begin{theorem}\label{digital-2}
In the $n$-period CRR binomial model, the binomial error $\mbox{Err}_{CRR}(n)$ for the prices of European digital call options with barrier $L>K$ is:
\begin{itemize}
\item
for a down-and-out digital call option:
\begin{align*}
\mbox{Err}(n)&=e^{-rT}\Biggl[(\tilde{E}_1+\tilde{E}_2\Delta^L_n)\frac{1}{\sqrt{n}}\\
&+(\tilde{F}_1+\tilde{F_2}\Delta^L_n+\tilde{F}_3(\Delta^L_n)^2\frac{1}{n}\Biggl]+O\Biggl(\frac{1}{n^{3/2}}\Biggl);\\
\end{align*}
\item
for a down-and-in digital call option:
\begin{align*}
\mbox{Err}(n)&=e^{-rT}\Biggl[(\tilde{G}_1+\tilde{G}_2\Delta^K_n+\tilde{G}_3\Delta^L_n)\frac{1}{\sqrt{n}}\\
&+(\tilde{H}_1+\tilde{H}_2(\Delta^K_n)^2+\tilde{H}_3\Delta^L_n+\tilde{H}_4(\Delta^L_n)^2)\frac{1}{n}\Biggl]+O\Biggl(\frac{1}{n^{3/2}}\Biggl),\\
\end{align*}
\end{itemize}
The list of the constants in postponed in Appendix A.
\end{theorem}

\textit{Proof.} By proceeding similarly to the proof of Theorem \ref{digital-1}, we need here to find an asymptotic expansion of the binomial price for a down-and-out digital call found in Proposition \ref{prop-2}.
In order to do this we apply (5.7), (5.11) and (5.15) in \textcolor{blue}{Lin\&Palmer} (\textcolor{blue}{2013}).
The down-and-in case is then obtained by considering the difference of the asymptotic expansion for the vanilla digital call and the down-and-out digital call.
\cvd

\begin{remark}
The term $\Delta^K_n$ does not appear in the expansion for the down-and-out option in Theorem \ref{digital-2}. The intuitive reason is that in this case $L>K$ and since the option stays alive if the stock price is above $L$, and therefore above $K$, the position of $K$ has no influence.
\end{remark}

\begin{remark}
In the error expansions found in Theorem \ref{digital-2} for down-and-in and down-and-out digital call options with $L>K$ it is not possible to totally vanish the contribution of order $\frac{1}{\sqrt{n}}$ by setting $\Delta^L_n=0$ and $\Delta^K_n=0$. In fact there is a constant term of order $\frac{1}{\sqrt{n}}$ that can't be nullified. A possibility in order to get an algorithm of order $\frac{1}{n}$ is to set $L$ and $K$ such that $\Delta^L_n=0=\Delta^K_n$ and then explicitly calculate the constant coefficient that multiplies $\frac{1}{\sqrt{n}}$ in order to subtract it to the binomial approximated price.
\end{remark}

Theorem \ref{digital-1} and Theorem \ref{digital-2} suggest us how to set the barrier $L$ and the strike $K$ in the binomial tree scheme in order to get an algorithm of order $\frac{1}{n}$. This theoretical result is enhanced by the numerical examples presented in Section \ref{sect-numerics}.

 Unfortunately, the extension of the previous reasoning for digital double barrier options is not straightforward because no manageable closed-form formulas of the CRR binomial prices exist in this case.\\
A possibility to deal with this issue is to use a completely different approach. We tried to extend to discontinuous payoffs the theoretical result in \textcolor{blue}{Gobet} (\textcolor{blue}{2001}), that studies the CRR binomial approximation error for double barrier options on a generic continuous payoff function by using PDE techniques. We expected to obtain that the contribution of order $\frac{1}{\sqrt{n}}$ could be explicitly written as dependent on two different sources: the position of the barriers and the position of the discontinuity point with respect to the nodes of the tree. Currently we are able to give just an upper bound of the CRR binomial approximation error in this more complex case, however we address it to a future work.

\section{Numerical results}\label{sect-numerics}
\begin{figure}[H] \centering
\includegraphics[height=7cm]{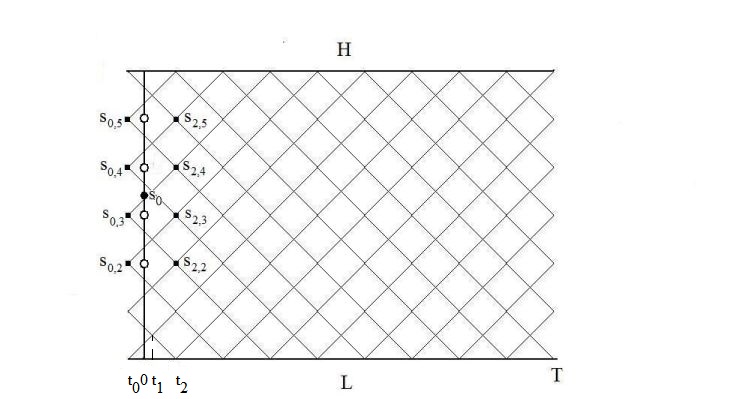}
\caption{\em Binomial Interpolated Lattice mesh for double barrier options.}
\label{intr-griglia}
\end{figure}

Theorem \ref{digital-1} and Theorem \ref{digital-2} on the asymptotic expansion of the CRR binomial approximation error, suggest that an algorithm of order $\frac{1}{n}$ can be obtained if in the $\log$-space the lower barrier $L$ lies exactly on a node of the tree (i.e. $\Delta^L_n=0$) and the strike $K$ is positioned halfway between two nodes at maturity (i.e. $\Delta^K_n=0$).\\
\mbox{\quad}To this end, we adapt here the Binomial Interpolated Lattice introduced in \textcolor{blue}{Appolloni \textit{et al.}} (\textcolor{blue}{2013}) (BIL) for pricing digital options with a single barrier. The BIL procedure is an efficient algorithm for the pricing of double barrier call and put options that we now briefly recall. The idea is to define the time step $\Delta t$ of the algorithm such that in the $\log$-space the lower barrier $L$ and the higher barrier $H$ coincide exactly with two nodes of the tree at maturity. Then, if $\Delta \tau=\frac{T}{n}$ is the standard time step of a CRR tree with $n$ steps, one needs to set
\begin{equation}\label{new-time-step}
\Delta t=\Big(\frac{h-l}{2k\sigma}\Big)^2
\end{equation}
where 
\begin{equation}\label{kappa}
k=\Biggl\lceil\frac{h-l}{2 \sigma \sqrt{\Delta \tau}}\Biggl\rceil
\end{equation}
and
$$
h=\log H \quad \mbox{and} \quad l=\log L.
$$
We recall here that $\lceil x \rceil$, for every $x \in \mathbb{R}$, denotes the smallest integer not less than $x$.
Then the time step $\Delta t$ defined in (\ref{new-time-step}) is obliged to take some specific values in order to match both $L$ and $H$ and this implies that $\frac{T}{\Delta t} \notin \mathbb{N}$. In order to arrive ``close to'' time $0$, one needs to add two further steps of length $\Delta t$ in order to get a fictitious time $t_0<0$ and a time $t_1>0$ (see Figure \ref{intr-griglia}) so that the number of time steps of the procedure is set as $n^{'}=\lfloor \frac{T}{\Delta T}\rfloor +2$.

Since we do not know a priori if the initial price $s_0$ is a point of the lattice (and in general it is not), the approximating option price at $(0, s_0)$ is provided by suitable interpolations in time and in space involving the prices, which are computed by the standard backward induction, at times $t_0$ and $t_2$. To be precise we choose in $t_0$ and $t_2$ the two points below and the two above $s_0$. The price at $s_0$ is obtained by a Lagrange four points interpolation in space of the prices denoted in Figure \ref{intr-griglia} with the empty circles, such prices are obtained by a linear interpolation in time of the prices at the nodes denoted by squares. For more details one can refer to \textcolor{blue}{Appolloni \textit{et al.}} (\textcolor{blue}{2013}).

\mbox{\quad}Then, we can easily adapt the BIL algorithm to the case of digital options with a single barrier.  
In fact we just need to modify the choice of $k$ defined in (\ref{kappa}) and the time step $\Delta t$ defined in (\ref{new-time-step}) such that the barrier $L$ is a node of the tree (in particular we set it as a node at maturity) and the strike $K$ lies halfway between two nodes at maturity (i.e. it is a node from the penultimate period). We set $\tilde{k}=\log K$ and then we define the integer $k$ as follows
\begin{equation}\label{k-digital}
k = \Biggl\lceil \frac{\tilde{k}-l}{2\sigma\sqrt{\Delta \tau}}\Biggl\rceil+\frac{1}{2},
\end{equation}
so that the time step $\Delta t$ is now given by
\begin{equation}\label{delta-digital}
\Delta t = \Biggl(\frac{\tilde{k}-l}{2\sigma k}\Biggl)^2.
\end{equation}
The number of time steps of this adjusted procedure, that we call ``Adjusted BIL'', is set as $n^{'}=\lfloor \frac{T}{\Delta t}\rfloor +2$. In fact the price at time $0$ is then obtained by a backward induction and by proceeding through interpolations in time and in space involving some specified prices at times $t_0=0$ and $t_2=2\Delta t$ as for the BIL algorithm. 

\begin{remark}
By using the choice of $k$ as in (\ref{k-digital}) so that the time step $\Delta t$ is defined as in (\ref{delta-digital}) we are able to construct a mesh in the $\log$-space in which the barrier $L$ is a node at maturity and the strike $K$ is a node from the penultimate period. But we stress here that in the Adjusted BIL algorithm it is not enough to build a binomial mesh between $L$ and $K$, but we need to extend it above $K$ and this is straightforward. This is a structural difference with what done in the BIL algorithm for pricing double barrier options: in this case we just need to setup the mesh between the barriers $L$ and $H$. 
\end{remark}

\mbox{\quad}We now present some numerical results in order to compare the prices for single barrier digital options obtained with the standard CRR algorithm and those obtained with the Adjusted BIL algorithm. In particular we study down-and-out digital call options in two cases: first, when $L<K$ (Section \ref{prima}); secondly, when $L>K$ (Section \ref{seconda}). The down-and-in case provides similar results, so we omit it.

\subsection{Down-and-out digital call option with $L<K$}\label{prima}

\begin{figure}[H] \centering
\includegraphics[height=8cm]{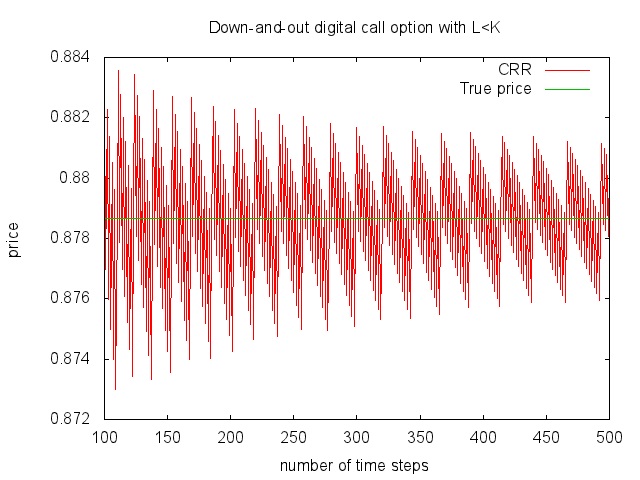}
\caption{\em CRR binomial approximation. Knock-out digital call option with $L=60, K=100, s_0=150, r=0.1, \sigma=0.25$ and $T=1$.}
\label{figA}
\end{figure}

We consider a down-and-out digital call option with lower barrier $L=60$, strike $K=100$ and initial stock value equal to $s_0=150$. The other parameters are: $r=0.1$, $\sigma=0.25$ and $T=1$. In Figure \ref{figA} we plot the European prices obtained by using the CRR binomial approximation and the true price calculated with the Black and Scholes formula, i.e.
\begin{equation}\label{true1}
C^{BS}_{do-digital}(s_0,K,T,L,r,\sigma)=e^{-rT}\Biggl[\Phi(d_{12})-\Phi(d_{22})\Biggl(\frac{s_0}{L}\Biggl)^{1-\frac{2r}{\sigma^2}}\Biggl],
\end{equation}
where $d_{12}$ and $d_{22}$ are defined in Appendix A and $\Phi(\cdot)$ is the standard normal distribution function.
We observe that the binomial price oscillates widely around the true price and this is due both on the position of $L$ and also on the position of $K$ with respect to the nodes of the tree. It is clear that the rate of convergence of the algorithm is of order $\frac{1}{\sqrt{n}}$.

\begin{figure}[H] \centering
\includegraphics[height=8cm]{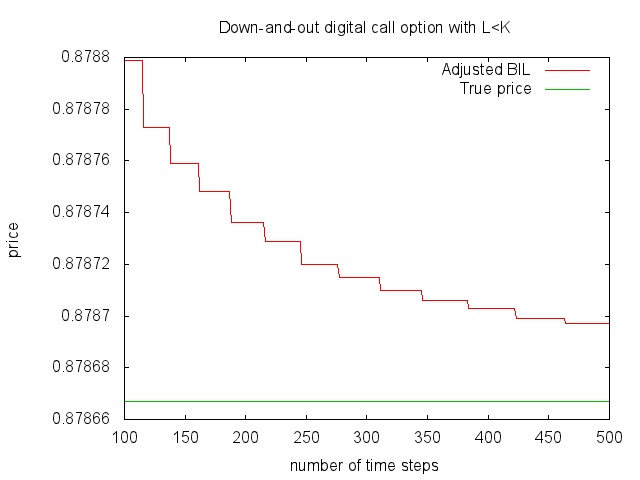}
\caption{\em Adjusted Binomial Interpolated Lattice. Knock-out digital call option with $L=60$, $K=100$, $s_0=150$,
  $r=0.1$, $\sigma=0.25$  and $T=1$. }
\label{figB}
\end{figure}

In Figure \ref{figB} we plot the prices obtained by using the Adjusted BIL algorithm. We remark here that in the $x$-axis we report the number $n$ of time steps corresponding to the CRR binomial approximation. In fact we recall that in the Adjusted BIL algorithm we define a new number of time steps $n^{'}$, different from $n$, but having the same order of magnitude. Now there are no oscillations and the convergence is of order $\frac{1}{n}$: the oscillations due $L$ and $K$ disappear and the convergence is monotone.

In Table \ref{tableA} we report the prices obtained with the CRR algorithm and the Adjusted BIL algorithm. In the first column we write the number $n$ of time steps of the CRR binomial approximation. The true price is calculated by using the Black and Scholes formula given in (\ref{true1}).

\begin{table}[H] \centering \scriptsize
{\begin{tabular} {@{}cccc@{}} \toprule && $L<K<s_0$ & \\  n & CRR & True& Adjusted BIL \\
\hline
100 &0.883147   & & 0.878791  \\
200 &0.879006   & & 0.878732  \\
400 &0.880340   & 0.878667 & 0.878700 \\
800 &0.876786   & & 0.878684\\
1600 &0.878863      &  & 0.878676 \\
3200 &0.877873    &  & 0.878671 \\ 
\hline
\end{tabular}}  \quad

\caption{\em Knock-out European digital call options prices with $L=60$, $K=100$, $s_0=150$,
  $r=0.1$, $\sigma=0.25$  and $T=1$.}
  \label{tableA}
\end{table}
\normalsize

\subsection{Down-and-out digital call option with $L>K$}\label{seconda}

\begin{figure}[H] \centering
\includegraphics[height=10cm]{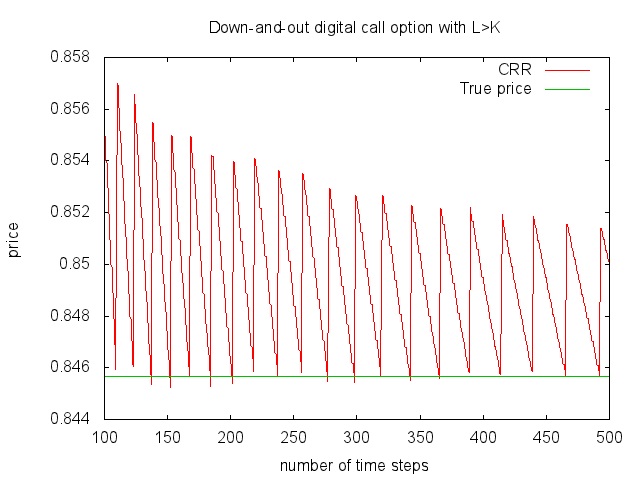}
\caption{\em CRR binomial approximation. Knock-out digital call option with $K=60$, $L=100$, $s_0=150$,
  $r=0.1$, $\sigma=0.25$  and $T=1$. }
\label{figC}
\end{figure}

We now consider a down-and-out digital call option with strike $K=60$, lower barrier $L=100$ and initial stock price $s_0=150$. The other parameters are: $r=0.1$, $\sigma=0.25$ and $T=1$. In Figure \ref{figC} we plot the European CRR binomial prices and the true price that is given by the following Black and Scholes formula
\begin{equation}\label{bs2}
C^{BS}_{do-digital}(s_0, K, T, L,r,\sigma) = e^{-rT}\Biggl[\Phi(d_{32})-\Biggl(\frac{s_0}{L}\Biggl)^{1-\frac{2r}{\sigma^2}}\Phi(d_{42})\Biggl],
\end{equation}
where $d_{32}$ and $d_{42}$ are defined in Appendix A.\\
We observe that in this case the oscillations are fewer than the case $L<K$ and this is due to the fact that the position of the strike $K$ has no influence in the error expansion. In fact the option stays alive when the stock price is above $L$ and therefore above $K$, so the position of $K$ with respect to the nodes of the tree has no influence. Then the term of order $\frac{1}{\sqrt{n}}$ is only due on the position of $L$, as remarked in Theorem \ref{digital-2}, and a constant term. However, from the numerical computations it turns out that the constant term is of order $10^{-3}$, so it really does not affect the error. But the convergence is still slow, i.e. the CRR algorithm has order $\frac{1}{\sqrt{n}}$.

\begin{figure}[H] \centering
\includegraphics[height=10cm]{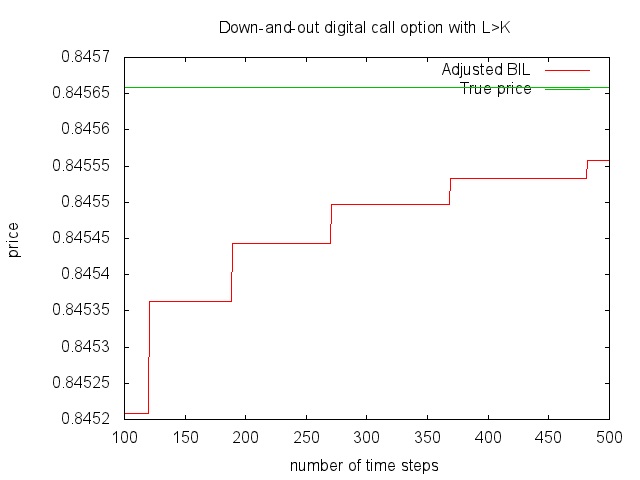}
\caption{\em Adjusted Binomial Interpolated Lattice. Knock-out digital call option with $K=60$, $L=100$, $s_0=150$,
  $r=0.1$, $\sigma=0.25$  and $T=1$. }
\label{figD}
\end{figure}

In Figure \ref{figD} we plot the European prices obtained with the Adjusted BIL algorithm and the Black and Scholes price given in (\ref{bs2}). We observe that here the convergence is monotone because the binomial mesh is constructed such that the lower barrier $L$ lies exactly on a layer of nodes. Then the procedure is of order $\frac{1}{n}$.

In Table \ref{tableB} we report the European prices of a down-and-out digital call option with lower barrier $L>K$ obtained with the CRR algorithm and the Adjusted BIL algorithm. As usual, $n$ denotes the number of time steps of the CRR binomial approximation. The true price is calculated by using the Black and Scholes formula given in (\ref{bs2}).

\begin{table}[H] \centering \scriptsize
{\begin{tabular} {@{}cccc@{}} \toprule && $K<L<s_0$ & \\  n & CRR & True& Adjusted BIL \\
\hline
100 &0.855913 &&0.844983 \\
200 &0.846415 &&0.845304 \\
400 &0.849497 && 0.845484 \\
800 &0.846188 &0.845659& 0.845571 \\
1600& 0.846107 && 0.845615 \\
3200& 0.846252 &&0.845637 \\ 
\hline
\end{tabular}}  \quad
\caption{\em Knock-out European digital call options prices with $K=60$, $L=100$, $s_0=150$,
  $r=0.1$, $\sigma=0.25$  and $T=1$.}
  \label{tableB}
\end{table}
\normalsize

\begin{remark}
The theoretical proof that the rate of convergence of the Adjusted BIL algorithm is of order $\frac{1}{n}$ is a direct consequence of Proposition 1 in \textcolor{blue}{Appolloni \textit{et al.}} (\textcolor{blue}{2013}), Theorem \ref{digital-1} and Theorem \ref{digital-2}. In fact, in the Adjusted BIL algorithm we get the price in $(0,s_0)$ by suitable interpolations of some selected CRR prices at times $t_0$ and $t_2$, as in the standard version of the BIL algorithm (see Figure \ref{intr-griglia}). But the interpolation rule preserves the error committed by approximating the continuous-time prices at the selected nodes by the CRR ones. So if the CRR binomial approximation error at these nodes is of order $\frac{1}{n}$, than the Adjusted BIL algorithm is still of order $\frac{1}{n}$.  
\end{remark}

\section{Conclusions}
We give here an explicit asymptotic expansion of the approximation error related to the standard CRR tree for pricing digital options with a single barrier. The theoretical results suggest us how to set a binomial algorithm such that the worst contribution term in the error expansion, that is of order $\frac{1}{\sqrt{n}}$ (where $n \in \mathbb{N}$ is the number of time steps of the algorithm), is nullified. We get an efficient lattice procedure enhanced by numerical examples.\\
\mbox{\quad} The extension of the reasoning to the case of double barrier digital options is not straightforward. Our idea is to use a different approach, based on PDE techniques, in order to study theoretically the CRR binomial approximation error. We found an upper bound of the error, but the result is still partial, so we address it to a future work.\\

\textbf{Acknowledgments} The authors thank Professor L. Caramellino and Professor A. Zanette for the useful comments and their valuable assistance and Professor K. Palmer for participating in discussions.

\section*{Appendix A}
We report here the list of the constants that appear in Theorem \ref{digital-1} and Theorem \ref{digital-2}:
\begin{align*}
d_{11}&=\frac{\log \frac{s_0}{K}+(r+\frac{1}{2}\sigma^2)T}{\sigma\sqrt{T}}, \quad d_{12}=d_{11}-\sigma\sqrt{T},\\
d_{21}&=\frac{\log \frac{L^2}{s_0K}+(r+\frac{1}{2}\sigma^2)T}{\sigma\sqrt{T}}, \quad d_{22}=d_{21}-\sigma\sqrt{T},\\
d_{31}&=\frac{\log \frac{s_0}{L}+(r+\frac{1}{2}\sigma^2)T}{\sigma\sqrt{T}}, \quad d_{32}=d_{31}-\sigma\sqrt{T},\\
d_{41}&=\frac{\log \frac{L}{s_0}+(r+\frac{1}{2}\sigma^2)T}{\sigma\sqrt{T}}, \quad d_{42}=d_{41}-\sigma\sqrt{T},\\
\alpha&=\frac{r-\frac{1}{2}\sigma^2}{2\sigma}, \quad \hat{\alpha}=\alpha+\frac{\sigma}{2},\\
\beta&=\frac{\sigma^4-4\sigma^2r+12r^2}{48\sigma}, \quad \hat{\beta}=-\beta-\frac{\sigma r}{6},\\
\hat{g}_i&=2T(\hat{\alpha}^2d_{i1}+\hat{\beta}\sqrt{T})+\Biggl(\frac{2\hat{\alpha}\sqrt{T}}{3}-\frac{d_{i1}}{12}\Biggl)(1-d^2_{i1}),i=1,2,3,4\\
g_i&=2T(\hat{\alpha}^2d_{i2}+\hat{\beta}\sqrt{T})+\Biggl(\frac{2\hat{\alpha}\sqrt{T}}{3}-\frac{d_{i2}}{12}\Biggl)(1-d^2_{i2}), i=1,2,3,4\\
G_1&=\frac{s_0}{\sqrt{2\pi}}e^{-\frac{d^2_{11}}{2}}(\hat{g_1}-g_1),\, G_2=\frac{s_0}{\sqrt{2\pi}}\Biggl(\frac{s_0}{L}\Biggl)^{-1-\frac{d^2_{21}}{2}}e^{-\frac{d^2_{21}}{2}}(\hat{g_2}-g_2),\\
G_3 &= \frac{s_0}{\sqrt{2\pi}}e^{-\frac{d^2_{31}}{2}}(\hat{g_3}-\frac{K}{L}g_3),\, G_4=\frac{s_0}{\sqrt{2\pi}}e^{-\frac{d^2_{31}}{2}}(\hat{g_4}-\frac{K}{L}g_4),\\
A_1&=4\sqrt{T}h_1(d_{21},d_{22}), \quad A_2=4\sqrt{T}h_1(d_{41},d_{42})+\frac{2x_0}{\sqrt{2\pi}}e^{-\frac{d^2_{31}}{2}}\Biggl(1-\frac{K}{L}\Biggl),\\
A_3&=4\sqrt{T}(h_1(-d_{41},-d_{42})-h_1(-d_{21},-d_{22}))-\frac{2s_0}{\sqrt{2\pi}}e^{-\frac{d^2_{31}}{2}}\Biggl(1-\frac{K}{L}\Biggl),\\
&h_{i}(x,y)=\Biggl(\frac{s_0}{L}\Biggl)^{-\frac{2r}{\sigma^2}}\Biggl(D\Biggl(\frac{r+\frac{\sigma^2}{2}}{2\sigma}\Biggl)^i\Phi(x)-\frac{s_0Ke^{-rT}}{L}\Biggl(\frac{r-\frac{\sigma^2}{2}}{2\sigma}\Biggl)^i\Phi(y)\Biggl),\\
&\mbox{for} \quad i=0,1,2,\\
B_1&=G_1-G_2+Ih_0(d_{21},d_{22}), B_2=B_1-G_1,\\
B_3&=G_3-G_4+Ih_0(d_{41},d_{42}), B_4=B_3-G_1,\\
B_5&=G_2+G_3-G_4+Ih_0(-d_{21},-d_{22})-Ih_0(-d_{41},-d_{42}), B_6=B_5-G_1,\\
I&=\Biggl(\frac{4\beta+\frac{16}{3}\alpha^3}{\sigma}\Biggl)\log\Biggl(\frac{s_0}{L}\Biggl)T,\\
C_1&=\frac{2s_0}{\sqrt{2\pi}}\Biggl(\frac{s_0}{L}\Biggl)^{-1-\frac{2r}{\sigma^2}}e^{-\frac{d^2_{21}}{2}}\sigma\sqrt{T}, \,
\end{align*}
\begin{align*}
C_2&=\frac{s_0}{\sqrt{2\pi}}e^{-\frac{d^2_{31}}{2}}\Biggl(d_{31}-\frac{K}{L}d_{32}\Biggl),\\
C_3&=\frac{s_0}{\sqrt{2\pi}}e^{-\frac{d^2_{31}}{2}}\Biggl(d_{41}-\frac{K}{L}d_{42}\Biggl), \, C=\frac{1}{2}(C_2-C_3),\\
D_1&=\frac{s_0}{2\sqrt{2\pi}}e^{-\frac{d^2_{11}}{2}}\sigma\sqrt{T}-\frac{C_1}{4},\, D_2=\frac{C_1}{4}, D_3=D_1+D_2,\\
E_1&=8Th_2(d_{21},d_{22})+C_1, E_2=8Th_2(d_{41},d_{42})+\frac{1}{2}(3C_2+C_3),\\
E_3&=8T(h_2(-d_{21},-d_{22})-h_2(-d_{41},-d_{42}))-C_1+\frac{1}{2}(3C_2+C_3),\\
\tilde{A}_1&=\Biggl(\frac{s_0}{L}\Biggl)^{1-\frac{2r}{\sigma^2}}\frac{e^{-\frac{d^2_{22}}{2}}}{\sqrt{2\pi}},\\
\tilde{A}_2&=-2\tilde{A}_1-4\alpha\sqrt{T}\Phi(d_{22})\Biggl(\frac{s_0}{L}\Biggl)^{1-\frac{2r}{\sigma^2}},\\
\tilde{B}_1&=\Biggl(\frac{s_0}{L}\Biggl)^{1-\frac{2r}{\sigma^2}}\Biggl[g_2\frac{e^{-\frac{d^2_{22}}{2}}}{\sqrt{2\pi}}-I\Phi(d_{22})\Biggl],\\
\tilde{B}_2&=\Biggl(\frac{s_0}{L}\Biggl)^{1-\frac{2r}{\sigma^2}}\Biggl(-\frac{d_{22}}{2}\Biggl)\frac{e^{-\frac{d^2_{22}}{2}}}{\sqrt{2\pi}},\\
\tilde{B}_3&=\Biggl(\frac{s_0}{L}\Biggl)^{1-\frac{2r}{\sigma^2}}\frac{e^{-\frac{d^2_{22}}{2}}}{\sqrt{2\pi}}[2d_{22}-4\alpha\sqrt{T}],\\
\tilde{B}_4&=\Biggl(\frac{s_0}{L}\Biggl)^{1-\frac{2r}{\sigma^2}}\Biggl[\frac{e^{-\frac{d^2_{22}}{2}}}{\sqrt{2\pi}}(-2d_{22}+8\alpha\sqrt{T})+8\alpha^2T\Phi(d_{22})\Biggl],\\
c_1&=\frac{e^{-\frac{d^2_{12}}{2}}}{\sqrt{2\pi}}, \, c_2=-\frac{d_{12}}{2}\frac{e^{-\frac{d^2_{12}}{2}}}{\sqrt{2\pi}},\\
\tilde{c}&=\frac{d^3_{11}+d_{11}d^2_{12}+2d_{12}-4d_{11}}{24}+\frac{(2-d_{11}d_{12}-d^2_{11})\sqrt{T}}{6\sigma}r+\frac{Td_{11}}{2\sigma^2}r^2,\\
c_3&=\tilde{c}\frac{e^{-\frac{d^2_{12}}{2}}}{\sqrt{2\pi}},\\
\tilde{C}_1&=c_1-\tilde{A}_1, \, \tilde{C}_2=-\tilde{A}_2,\\
\tilde{D}_1&=c_2-\tilde{B}_1, \, \tilde{D}_2=c_3-\tilde{B}_2,\\
\tilde{D}_3&=-\tilde{B}_3, \, \tilde{D}_4=-\tilde{B}_4,\\
\tilde{E}_1&=-\epsilon_n\frac{e^{-\frac{d^2_{32}}{2}}}{\sqrt{2\pi}}+\Biggl(\frac{s_0}{L}\Biggl)^{1-\frac{2r}{\sigma^2}}\frac{e^{-\frac{d^2_{42}}{2}}}{\sqrt{2\pi}},\\
\end{align*}
\begin{align*}
\tilde{E}_2&=\frac{e^{-\frac{d^2_{32}}{2}}}{\sqrt{2\pi}}+\Biggl(\frac{s_0}{L}\Biggl)^{1-\frac{2r}{\sigma^2}}\Biggl(\frac{e^{-\frac{d^2_{42}}{2}}}{\sqrt{2\pi}}+4\alpha\sqrt\Phi(d_{42})\Biggl),\\
\tilde{F}_1&=\frac{e^{-\frac{d^2_{32}}{2}}}{\sqrt{2\pi}}\Biggl(g_3-\frac{d_{32}}{2}\epsilon_n^2\Biggl)+\Biggl(\frac{s_0}{L}\Biggl)^{1-\frac{2r}{\sigma^2}}\frac{e^{-\frac{d^2_{42}}{2}}}{\sqrt{2\pi}}\Biggl(\frac{d_{42}}{2}\epsilon_n^2-g_4\Biggl)\\
&+\Phi(d_{42})I\Biggl(\frac{s_0}{L}\Biggl)^{1-\frac{2r}{\sigma^2}},\\
\tilde{F}_2&=\frac{e^{-\frac{d^2_{32}}{2}}}{\sqrt{2\pi}}d_{32}\epsilon_n+\Biggl(\frac{s_0}{L}\Biggl)^{1-\frac{2r}{\sigma^2}}\Biggl(\frac{e^{-\frac{d^2_{42}}{2}}}{\sqrt{2\pi}}\epsilon_nd_{42}-4\epsilon_n\alpha\sqrt{T}\Biggl),\\
\tilde{F}_3&=-\frac{d_{32}}{2}\frac{e^{-\frac{d^2_{32}}{2}}}{\sqrt{2\pi}}+\Biggl(\frac{s_0}{L}\Biggl)^{1-\frac{2r}{\sigma^2}}\frac{e^{-\frac{d^2_{42}}{2}}}{\sqrt{2\pi}}\Biggl(\frac{d_{42}}{2}-4\alpha\sqrt{T}\Biggl)\\
&-\Biggl(\frac{s_0}{L}\Biggl)^{1-\frac{2r}{\sigma^2}}\Phi(d_{42})8\alpha^2T,\\
\tilde{G}_1&=-\tilde{E}_1, \, \tilde{G}_2=c_1,
\tilde{G}_3=-\tilde{E}_2,\\
\tilde{H}_1&=c_2-\tilde{F}_1,\,
\tilde{H}_2=c_3,\,
\tilde{H}_3=-\tilde{F}_2,\,
\tilde{H}_4=-\tilde{F}_3.\\
\end{align*}

\end{document}